\documentclass[journal=aamick,manuscript=article,layout=traditional]{achemso}

\usepackage{natbib}
\usepackage{graphicx}

\title{Ionic liquids under confinement: From systematic variations of the ion and pore sizes towards an understanding of structure and dynamics in complex porous carbons}

\author{El Hassane Lahrar}
\affiliation[1]{CIRIMAT, Universit\'e de Toulouse, CNRS, B\^at. CIRIMAT, 118, route de Narbonne 31062 Toulouse cedex 9, France}
\affiliation[2]{R\'eseau sur le Stockage \'Electrochimique de l'\'Energie (RS2E), F\'ed\'eration de Recherche CNRS 3459, HUB de l'\'Energie, Rue Baudelocque,  80039 Amiens, France}
\author{Anouar Belhboub}
\affiliation[1]{CIRIMAT, Universit\'e de Toulouse, CNRS, B\^at. CIRIMAT, 118, route de Narbonne 31062 Toulouse cedex 9, France}
\affiliation[2]{R\'eseau sur le Stockage \'Electrochimique de l'\'Energie (RS2E), F\'ed\'eration de Recherche CNRS 3459, HUB de l'\'Energie, Rue Baudelocque,  80039 Amiens, France}
\author{Patrice Simon}
\affiliation[1]{CIRIMAT, Universit\'e de Toulouse, CNRS, B\^at. CIRIMAT, 118, route de Narbonne 31062 Toulouse cedex 9, France}
\affiliation[2]{R\'eseau sur le Stockage \'Electrochimique de l'\'Energie (RS2E), F\'ed\'eration de Recherche CNRS 3459, HUB de l'\'Energie, Rue Baudelocque,  80039 Amiens, France}
\author{C\'eline Merlet}
\affiliation[1]{CIRIMAT, Universit\'e de Toulouse, CNRS, B\^at. CIRIMAT, 118, route de Narbonne 31062 Toulouse cedex 9, France}
\affiliation[2]{R\'eseau sur le Stockage \'Electrochimique de l'\'Energie (RS2E), F\'ed\'eration de Recherche CNRS 3459, HUB de l'\'Energie, Rue Baudelocque,  80039 Amiens, France}
\email{merlet@chimie.ups-tlse.fr}

\date{}

\begin{document}

\begin{tocentry}
\includegraphics[scale=1.0]{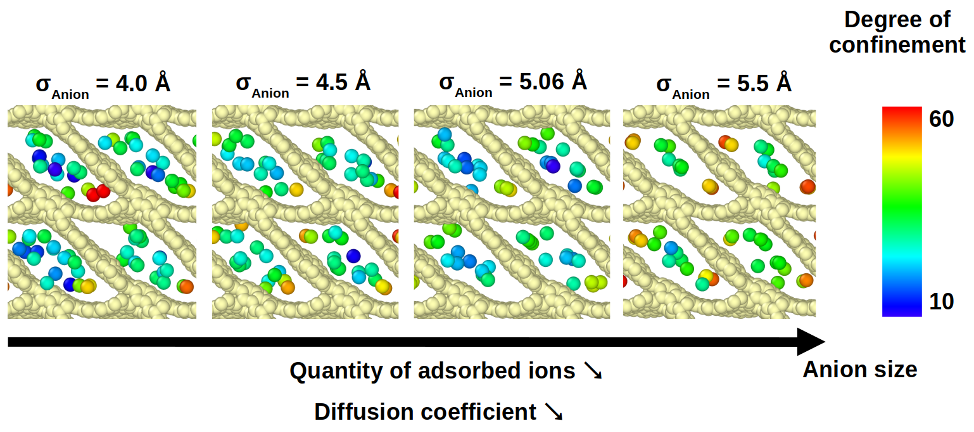}
\end{tocentry}

\begin{abstract}
We use molecular simulations of an ionic liquid in contact with a range of nanoporous carbons to investigate correlations between ion size, pore size, pore topology and properties of the adsorbed ions. We show that diffusion coefficients increase with the anion size and, surprisingly, with the quantity of adsorbed ions. Both findings are interpreted in terms of confinement: when the in-pore population increases additional ions are located in less confined sites and diffuse faster. Simulations in which the pores are enlarged while keeping the topology constant comfort these observations. The interpretation of properties across structures is more challenging. An interesting point is that smaller pores do not necessarily lead to larger confinement. In this work, the highest degrees of confinements are observed for intermediate pore sizes. We also show a correlation between the quantity of adsorbed ions and the ratio between the maximum pore diameter and the pore limiting diameter.
\end{abstract}

\section{Introduction}

Ionic liquids are considered as promising electrolytes for Electrochemical Double Layer Capacitors (EDLCs), also called supercapacitors, thanks to to their wide electrochemical windows, allowing for larger energy densities, and their higher safety compared to organic electrolytes~\citep{Zhong15,Beguin14,Brandt13,Lewandowski10}. One downside of using ionic liquids compared to organic or aqueous electrolytes is their relatively low ionic conductivity which can limit the power density of the devices. In EDLCs, charge storage occurs through reversible ion adsorption at the electrode/electrolyte interface. Porous carbons are widely used as electrode materials owing to their low cost and the large surface areas they provide~\citep{Simon13,Liu17,Fic18}. Understanding the relationships between the ions and carbon structures and the electrochemical performance of supercapacitors is a major challenge as the disordered nature of most of the carbons used commonly renders this interface very difficult to characterise. Moreover, pure ionic liquids are highly concentrated and the description of the electrostatic interactions in such systems is still a challenge~\citep{Lee17b,Lhermerout18}.

Recent advances in \emph{in situ} experimental methods, such as Nuclear Magnetic Resonance (NMR) and Electrochemical Quartz Crystal Microbalance (EQCM), have provided invaluable insights into the charge storage mechanisms of ionic liquid based supercapacitors~\citep{Tsai14,Forse16,Forse15}. When a potential difference is applied between the carbon electrodes, the charge storage can occur through counter-ion adsorption, co-ion desorption and ion exchange. It was demonstrated that the charging mechanism depends on the nature of the electrolyte and is usually different for the positive and negative electrodes due to the asymmetry between anions and cations. It was also shown that the total number of ions in the pores affects the diffusion coefficients of these adsorbed species~\cite{Forse17}, which therefore impacts the power density of the devices. While these techniques have allowed considerable progress in the understanding of the charging mechanisms, there is still no clear way to predict what the charging mechanisms will be when simply knowing the nature of the ions and the structure of the carbon. 

A number of theoretical models have been proposed in the past to calculate the number of ions adsorbed in a porous electrode at a given potential difference and predict the corresponding electrochemical performances~\citep{Kornyshev07,Huang08}. These models have the advantage of being very fast and have permitted the establishment of key concepts such as the superionic effect~\citep{Kondrat11}, i.e. the fact that ions of the same charge can be nearest neighbors in small pores thanks to the enhanced charge screening from the pore walls. In addition to theoretical models, molecular dynamics simulations have been used extensively to probe the structural, capacitive and dynamic properties of the interface between pure ionic liquids and porous electrodes~\citep{Merlet13c,Fedorov14}. While being more computationally expensive, molecular simulations have a great advantage in their ability to describe the complex nature of the ions and the porous structure in a much more accurate way than analytical models. The relationship observed experimentally between the total number of ions in the pores and the diffusion coefficients could be reproduced in a number of simulations~\citep{Burt16,He16}. Nevertheless, such studies are still rare and usually focusing on a single porous carbon structure so that a clear picture of the correlations between ion size, porous structure and interfacial properties is still missing.

In this work, we report on a methodical molecular dynamics simulations study of the confinement effects for a pure ionic liquid in contact with a set of nanoporous carbons. We start by systematically varying the anion size to assess the influence of this property on the structural and dynamical properties of the ionic liquid in the bulk and under confinement. We show that while the evolution of the quantity of adsorbed ions with anion size is intuitive, the variation of the diffusion coefficients is unforeseen. We interpret the observed trends in terms of degrees of confinement of the ions in the porous carbons. We then focus on two carbon structures, which we enlarge arbitrarily to investigate the effect of pore size while keeping the same topology, and show that the obtained results are in agreement with the trend observed for the anion size evolution. Finally, we compare structural and dynamical properties across carbons with different topologies, a much more challenging task. We show that the confinement does not necessarily decrease with an increase of pore size and that the non regularity of the structure, tentatively assessed here through the ratio between the maximum pore diameter and the pore limiting diameter, seems to be correlated with the total number of ions adsorbing in the porous volume.

\section{Methods}

\subsection{Systems studied}

\begin{figure}[ht!]
\centering
\includegraphics[scale=0.2]{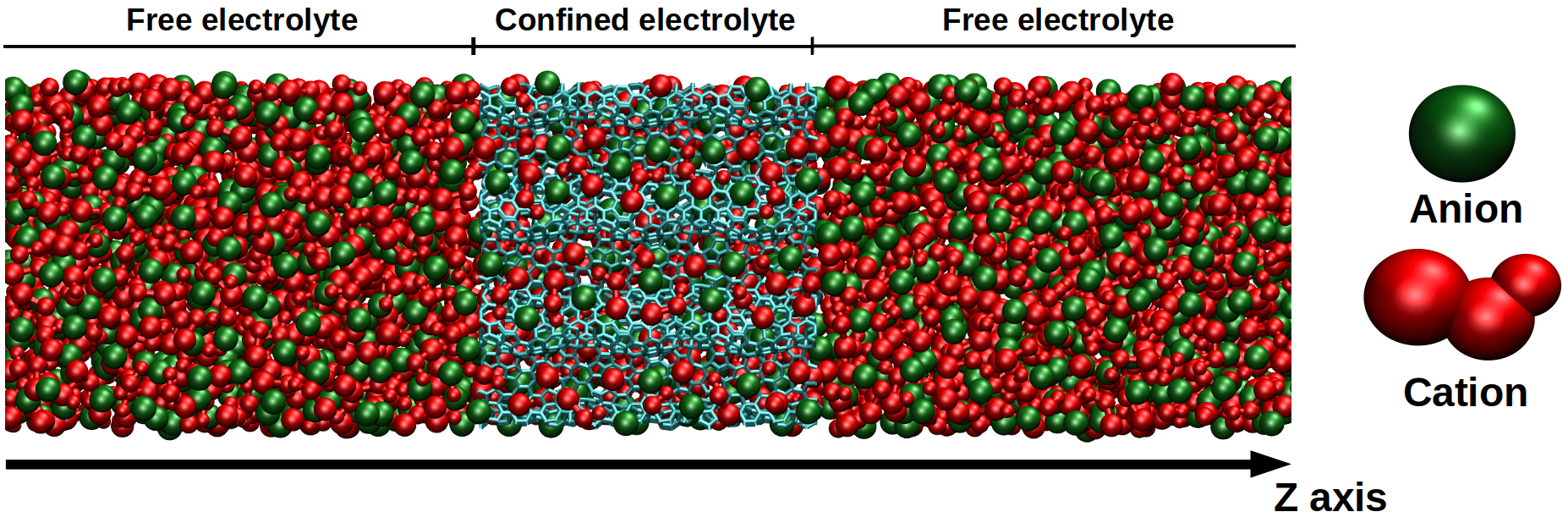}
\caption{Snapshot of one of the simulated systems: a pure ionic liquid in contact with a porous carbon. Anions are represented in green, cations in red and carbon atoms in light blue. This snapshot was generated using the VMD software.\citep{VMD}}
\label{fig:setup}
\end{figure}

Molecular dynamics (MD) simulations of a pure ionic liquid (1-butyl-3-methylimidazolium hexafluorophosphate, [BMI][PF$_6$], and derivatives) both in the bulk and in contact with a porous carbon have been carried out (Figure~\ref{fig:setup}). The electrolyte is represented by a coarse-grained model with three sites for the cation and one site for the anion~\citep{Roy10b}. The cation geometry is kept rigid. The intermolecular interactions are calculated as the sum of a Lennard-Jones potential and coulombic interactions:
\begin{equation}
u_{ij}(r_{ij})=4\varepsilon_{ij}[(\frac{\sigma_{ij}}{r_{ij}})^{12}-(\frac{\sigma_{ij}}{r_{ij}})^6]+\frac{q_iq_j}{4\pi\varepsilon_0r_{ij}}
\end{equation}
where $r_{ij}$ is the distance between sites $i$ and~$j$ and $\varepsilon_0$ is the permittivity of free space. $\sigma_{ij}$ and $\epsilon_{ij}$ are the Lennard-Jones parameters defining respectively the position of the repulsive wall and the depth of the energy well (see Figure~S1). Crossed parameters are calculated by Lorentz-Berthelot mixing rules. The parameters for the ions are taken from the work of Roy and Maroncelli~\citep{Roy10b} and the ones for the carbon atoms from the article of Cole and Klein~\cite{Cole83}. To study the effect of the anion size on the properties of the systems, we simply vary the $\sigma$ parameter for the anion. Starting from the original model with $\sigma_{\rm Anion}=5.06$~\r{A}, we test three other values: 4.0~\r{A}, 4.5~\r{A} and 5.5~\r{A}. The evolution of the Lennard-Jones potential with $\sigma_{\rm Anion}$ is shown in Supporting Information.

For the simulations of the pure ionic liquid in contact with porous carbons, we study a set of 14 carbons with equal densities (1~g.cm$^{-3}$). The pore size distributions of all carbons are given in Figure~\ref{fig:carbons} along with snapshots of a few carbons. Snapshots for all the carbons are provided in Supplementary Information. Amongst the carbons studied, 10 are ordered with a well-defined pore size, or having at most a bi-modal distribution. These carbons usually have a channel oriented in the $z$ direction. The other 4 carbons are disordered and are characterised by a much wider pore size distribution. The ordered carbons were generated by Deringer~\emph{et al.}~\citep{Deringer18} through quench molecular dynamics using a machine-learning based force field developed using the Gaussian Approximated Potentials approach~\citep{Deringer17} and are thus designated as ``GAP" carbons. The disordered carbons are taken from the work of Palmer~\emph{et~al.}~\citep{Palmer10} and were also obtained from Quench Molecular Dynamics. For these carbons, we keep the original naming system of the authors, ``QMDNx" where ``Nx" is related to the quench rate: a higher number indicates a higher quench rate which usually corresponds to a more disordered carbon.

\begin{figure}[ht!]
\includegraphics[scale=0.23]{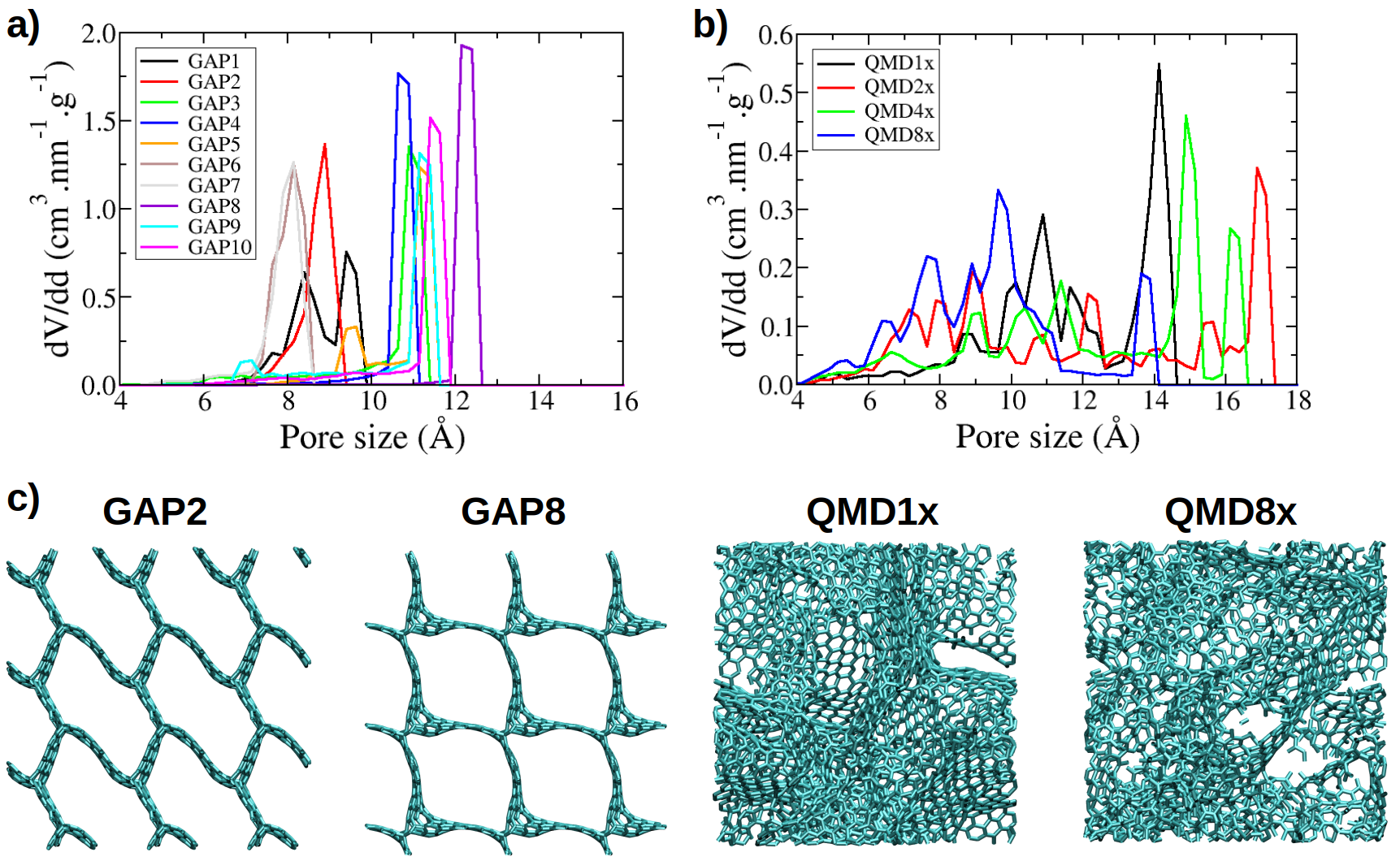}
\caption{Pore size distributions of different carbon structures studied in this work: a) ordered carbons~\citep{Deringer18}, b) disordered carbons~\citep{Palmer10}. Pore size distributions were obtained using the Poreblazer software~\citep{Sarkisov11}. Snapshots of a few selected structures, other structures can be seen in Supplementary Information.}
\label{fig:carbons}
\end{figure}

All simulations reported here have been conducted using the LAMMPS software~\citep{LAMMPS}. The timestep is set to 2~fs. Bulk simulations are done with cubic simulation boxes and contain 1200~ion pairs. The systems are first equilibrated in the NPT ensemble for 4~ns before collecting data for 10~ns in the NVT ensemble. The simulations with porous carbons also contain 1200 ion pairs. The simulation boxes have dimensions close to 48.5 \r{A} $\times$ 48.5 \r{A} $\times$ 210 \r{A} and depend on the anion size. The systems are first equilibrated in the NPT ensemble for 2 ns before collecting data for 10 ns in the NVT ensemble. The pressure of the NPT simulations is set to 1~atm and all simulations are done at 400~K. The barostat and thermostat time constants are 0.5~ps and 0.1~ps respectively. 

\subsection{Pair distribution functions and diffusion coefficients}

We characterise the structural and dynamical properties through pair distribution functions and diffusion coefficients. Pair distribution functions, or radial distribution functions, give a measure of the probability of finding a pair of atoms separated by a distance $r$, relative to the probability estimated for a completely random distribution at the same density. A possible expression for these pair distribution functions is:
 \begin{equation}
  g_{\alpha \beta}(||\mathbf{r_i}-\mathbf{r_j}||) = \frac{\rho_{\alpha\beta}^{(2)}(\mathbf{r_i},\mathbf{r_j})}{\rho_{\alpha}^{(1)}(\mathbf{r_i})\times\rho_{\beta}^{(1)}(\mathbf{r_j})},
 \end{equation}
where $\rho_{\alpha}^{(1)}$ and $\rho_{\alpha \beta}^{(2)}$ are respectively the one-body and two-body particle densities for ions of species $\alpha$ (and $\beta$).

Generally, in a uniform fluid, the diffusion is homogeneous along the three axes $x$, $y$ and~$z$; and the determination of the self-diffusion coefficients is done using the Einstein relation which relates this property to the mean-square displacement (MSD) of the molecules: 
 \begin{equation}
  D = \lim_{t\to\infty} \frac{1}{2dt} <| \Delta \mathbf{r_i}(t) |^2>,
 \end{equation}
where $d$ is the dimensionality of the system and $\Delta \mathbf{r_i}(t)$ is the displacement of a typical ion of the considered species in time $t$.

The presence of the porous carbon breaks the symmetry of the system, as shown in Fi\-gu\-re~\ref{fig:setup}, and we can define a region of ``free electrolyte" and a region of ``confined electrolyte". The determination of diffusion coefficients in such a system have been described in the literature~\citep{Liu04,Rotenberg07}. $D_{xx}$ and $D_{yy}$ are determined from the mean square displacements $<\Delta x^2(t)>$ and $<\Delta y^2(t)>$ of particles remaining in a given region. $P_i(t)$ is the survival probability for a particle in that given region :  
\begin{equation}
    D_{xx}(z_i) =  \lim\limits_{t \rightarrow \infty} \frac {<\Delta x_i^2(t)>} {2tP_i(t)},\hspace{0.5cm}   
    D_{yy}(z_i) =  \lim\limits_{t \rightarrow \infty} \frac {<\Delta y_i^2(t)>} {2tP_i(t)}
\end{equation}
The diffusion coefficient along the $z$ axis, $D_{zz}$, is determined from the autocorrelation of an eigenfunction based on the $z$ limit condition : 

\begin{equation}
D_{zz}(z_i)=-\left(\frac{L}{n\pi}\right)^2  \lim\limits_{t \rightarrow 0} \frac {\ln(<\psi_n^i(t) \psi_n^i(0)>)} {t},
\end{equation}
where $\psi_n^i(t)$ is given by:
\begin{equation}
\psi_n^i(z_i)= \sin\left(n\pi \frac{z(t)-z_{min}^i}{z_{max}^i-z_{min}^i}\right).
\end{equation}
In these equations, $z_{min}^i$ and $z_{max}^i$ are the coordinates that define the width of a given region $L = z_{max}^i - z_{min}^i $, and $n$ is an integer which should not affect the results much in the diffusive regime, here we choose $n=3$.

\section{Results and discussion}

\subsection{Effect of the anion size on the bulk properties}

We start our systematic study of the correlation between the ion size, the pore size and the properties of the electrolyte under confinement by looking at the effect of the ion size. Here, for the sake of fundamental understanding, we simply vary the $\sigma$ parameter of the anion in the Lennard-Jones potential employed. The coarse-grained model we use to represent the pure ionic liquid is well suited for this as the anion is described as a single spherical site~\citep{Roy10b}. While this is not realistic, molecular simulations give us this unique opportunity to assess the effect of such a variation. Starting from the original model of Roy and Maroncelli, developed to represent  [BMI][PF$_6$] and having a $\sigma_{\rm Anion}$ of 5.06~\r{A}, we investigate four different anion sizes: 4.0~\r{A}, 4.5~\r{A}, 5.06~\r{A} and 5.5~\r{A}. Since the variation of the $\sigma$ parameter for the anions modifies the intermolecular interactions, we need to start by characterising the structural and dynamical properties of the bulk liquid before turning to the characterisation of the same properties under confinement.

\begin{figure}[ht!]
\includegraphics[scale=0.26]{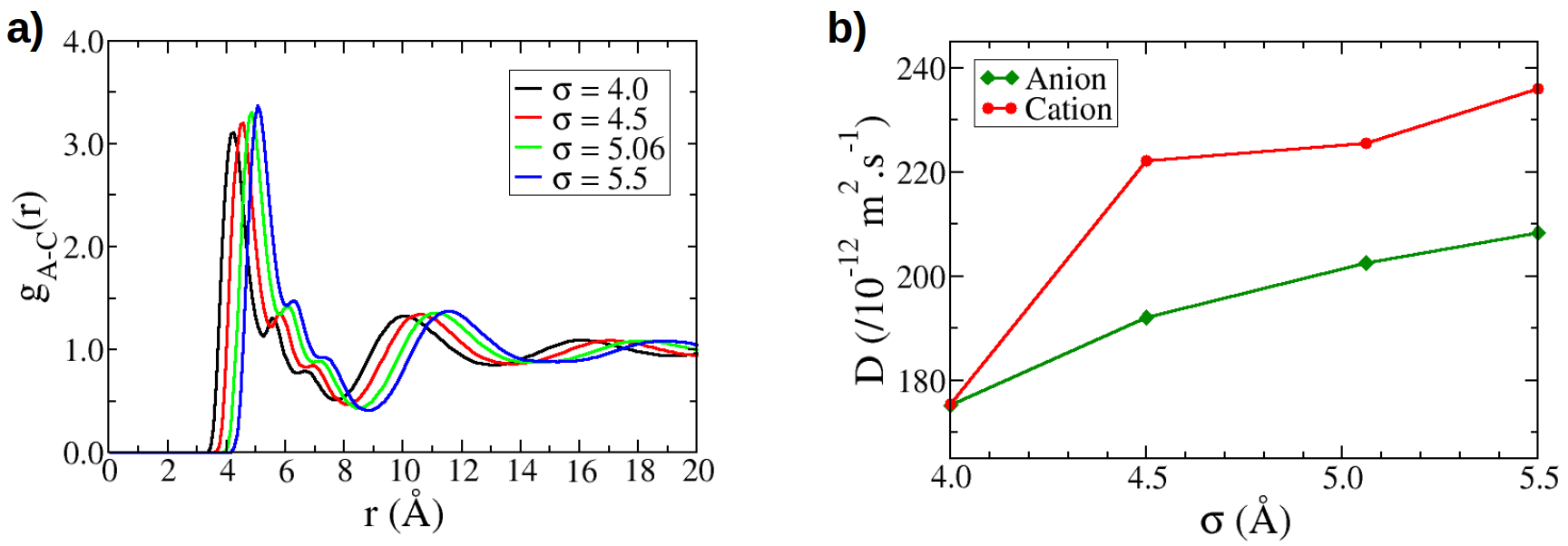}
\caption{a) Centers of mass radial distribution functions between anions and cations, and b) diffusion coefficients for bulk simulations of pure ionic liquids with various Lennard-Jones parameters ($\sigma$ for the anion.)}
\label{fig:bulkresults}
\end{figure}

Figure~\ref{fig:bulkresults}a gives the pair distribution functions between the centers of mass of the anions and the cations in the bulk. The curves for the different $\sigma_{\rm Anion}$ are similar but, as expected, the larger $\sigma_{\rm Anion}$ is, the more the first maximum is shifted to large distances. The heights of the first maximum and the following minimum also varies slightly with $\sigma_{\rm Anion}$. The average coordination number is constant with $\sigma_{\rm Anion}$ and close to 6. It is unclear at this point what drives the slightly more pronounced structuring for the larger $\sigma_{\rm Anion}$ but it is probably related to a different interplay between Lennard-Jones and electrostatic interactions for ions of different sizes. Another interesting point to note is the presence of two shoulders on the first peak of the pair distribution functions, these shoulders are due to different orientations of the cations with respect to the anion. 

Figure~\ref{fig:bulkresults}b gives the diffusion coefficients calculated for the bulk electrolyte. On this figure, we see that the diffusion coefficient increases when the anion size increases. Starting from the original $\sigma_{\rm Anion}$ of 5.06~\r{A}, we see a variation of up to ~20~\% in the diffusion coefficient. It is also interesting to note that while we modify only $\sigma_{\rm Anion}$, the diffusion coefficient of the cation is also affected. We believe that the fact that we obtain equal values for $\sigma_{\rm Anion}=4.0$~\r{A} is fortuitous. The increase in the diffusion coefficients can be related to the decrease in density when $\sigma_{\rm Anion}$ increases, densities have been known to affect diffusion~\citep{Rajput12,ChenluWang}. The increase in the ion-ion distance reducing the electrostatic interactions could also contribute to increasing the diffusion coefficients.

\subsection{Effect of the anion size on the confined properties}

\begin{figure}[ht!]
\includegraphics[scale=0.28]{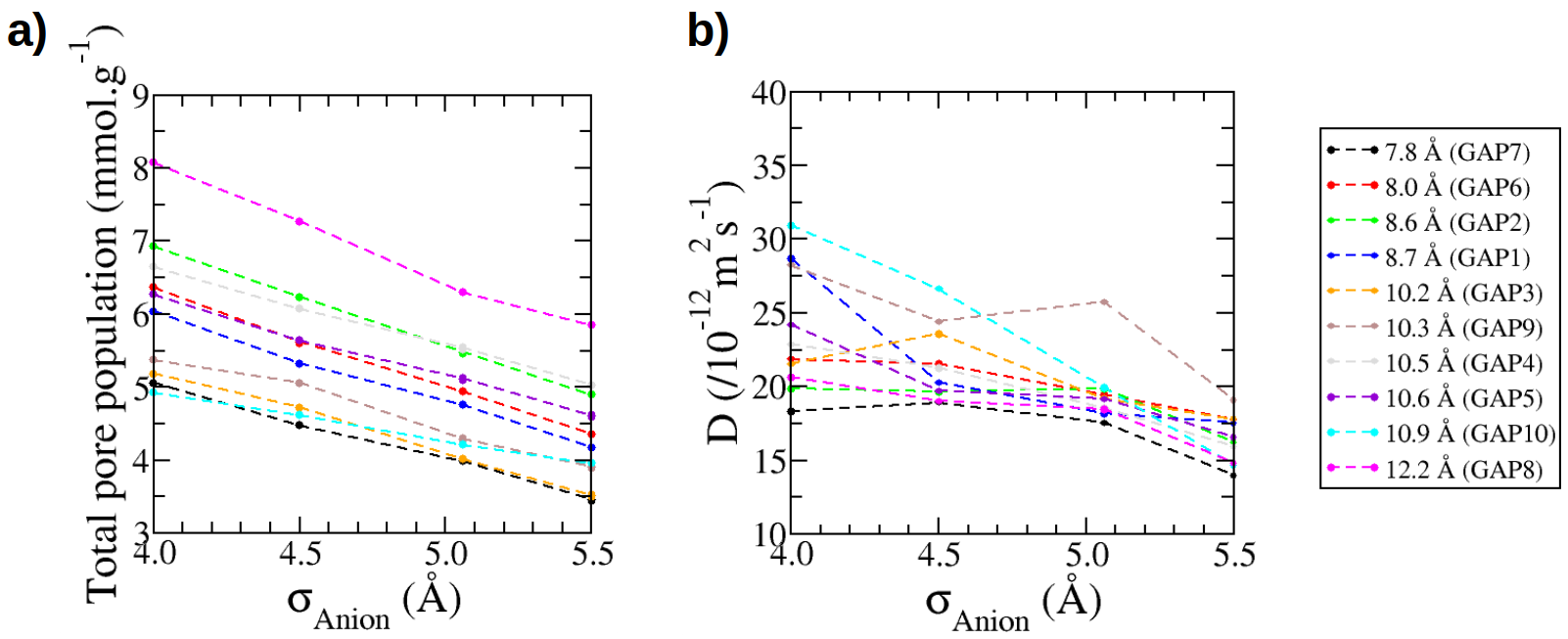}
\caption{a) Total pore populations, and b) diffusion coefficients for the anions confined in the GAP carbons with various Lennard-Jones parameters ($\sigma_{\rm Anion}$).}
\label{fig:TPP-Diff-GAPs-sigma}
\end{figure}

We now discuss the effect of the anion size on the confined properties of the electrolyte. In the remainder of this article, we focus mainly on the properties of the anions but similar conclusions can be drawn for the cations. In this part, we investigate ion adsorption and diffusion in 11 different carbon structures with average pore sizes comprised between 7.8~\r{A} and 12.2~\r{A}: 10 ordered carbons designated as GAPs and one disordered carbon designated as QMD4x. One important information to extract when looking at ion adsorption in porous materials is the quantity of ions actually adsorbed in the pores. This is relevant for energy storage as it can impact both the power density, i.e. how fast the systems can be charged or discharged, and the capacitance of such devices. The quantities of ions adsorbed in the porous carbons are given in Figure~\ref{fig:TPP-Diff-GAPs-sigma}. The total pore population is the sum of the numbers of anions and cations in the carbon normalised by the mass of the carbon material. We remind here that all carbons have the same density of 1~g.cm$^{-3}$. It is very clear from Figure~\ref{fig:TPP-Diff-GAPs-sigma}a that for a given carbon the variation of the total pore population with $\sigma_{\rm Anion}$ is monotonous. Not surprisingly, when the anion size increases, the quantity of ions in the pores decreases. It is interesting to note that the curves are almost linear and with similar slopes for all the carbons. This might be a feature specific to ordered carbons as the disordered carbon QMD4x does not show a linear trend (see Supplementary Information).

Figure~\ref{fig:TPP-Diff-GAPs-sigma}b gives the diffusion coefficients as a function of $\sigma_{\rm Anion}$ for the various GAP carbons. While the trend is not as clear as for total pore populations, most of the systems show a decrease in the diffusion when the anion size increases. It is thus the opposite of what was observed in the bulk simulation and it shows how big an effect the confinement has on the dynamical properties of the electrolyte. Also worth noticing is the fact that the diffusion coefficients of anions adsorbed in the carbons are around one order of magnitude smaller than in the bulk. This is a smaller decrease than the one observed in Pulse Field Gradient NMR experiments~\citep{Forse17} but it is in agreement with other simulation studies~\citep{Burt16,He16}.

From Figure~\ref{fig:TPP-Diff-GAPs-sigma}, it seems that the diffusion coefficients are larger when the total pore population is larger. This is even clearer in Figure~S4 in which we plot the diffusion coefficients as a function of the total pore population. This result is rather counter-intuitive and previous experiments~\citep{Forse17} and simulations~\citep{Burt16,He16} have shown a reverse trend. It is important to note however that previous correlations were done with a given electrolyte-carbon combination and the changes in populations were due to an applied potential difference. In the present case, the situation is very different because we are studying a set of electrolytes, due to the variation of the anion size, and a set of carbons, with different pore sizes and topologies.   

\begin{figure}[ht!]
\includegraphics[scale=0.22]{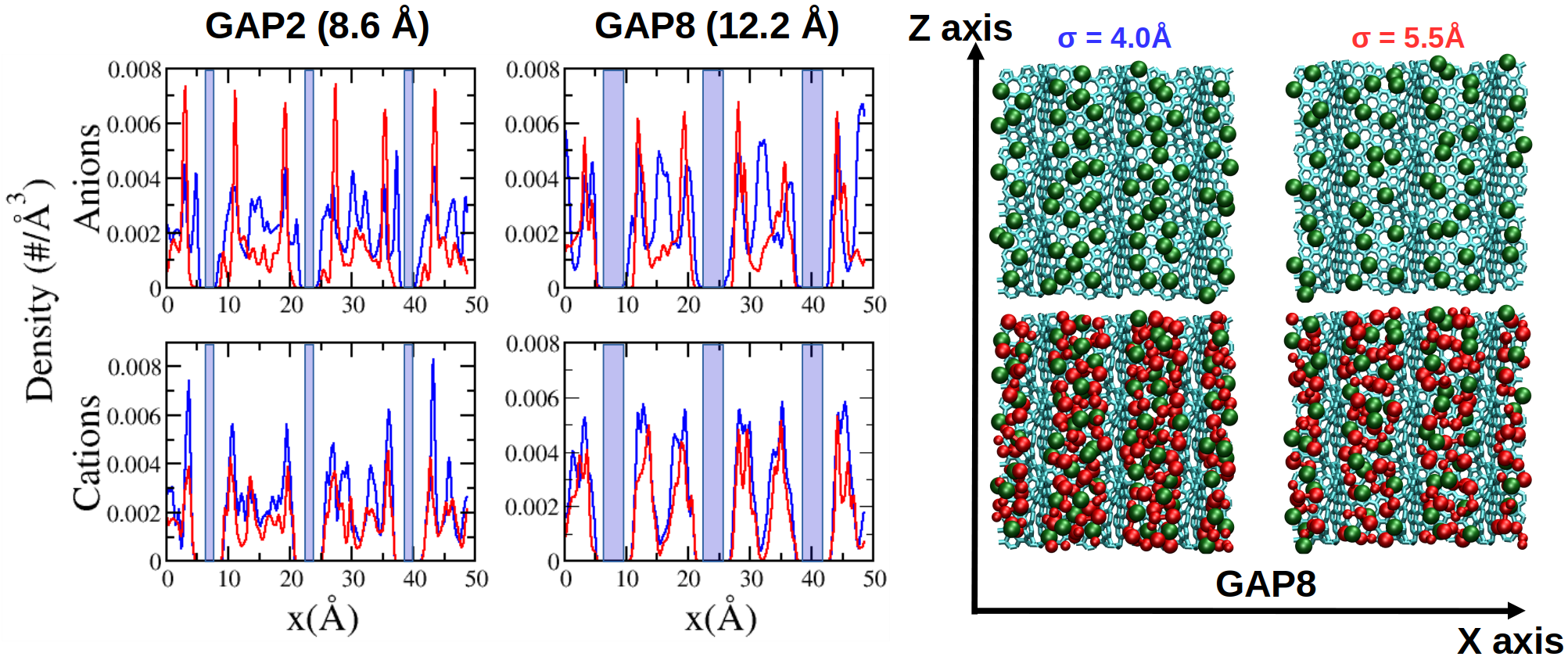}
\caption{Left: Ionic densities along the x-axis for the GAP2 and GAP8 carbons. Right: Snapshots showing the ions confined in the GAP8 carbon. Anions are represented in green, cations in red and carbon atoms in light blue. These snapshots were generated using the VMD software.~\citep{VMD}}
\label{fig:Densities-x}
\end{figure}

To explain the observations made on the correlation between the total pore population and the diffusion coefficients, we characterise the structure of the electrolyte in a finer way. First, we look at the ionic densities in the $x$ direction. The carbons we choose for this study have a topology in which the $x$ and $y$ axes more or less correspond to particular orientations in the carbons topology. The ionic densities as well as snapshots of the electrolyte-carbon system are shown in Figure~\ref{fig:Densities-x} for the smallest and the largest sizes of the anion. On this figure, it is again clear that a smaller anion size leads to a higher density of ions in the pores. More interestingly, and especially visible on the anion densities, the density seems to increase mostly in the center of the pore. In the case of GAP8, we even observe a change of structure from a bilayer of ions for large anions to a trilayer for small anions. Previous studies have shown that ions in the center of the pores move faster than ions in the first layer close to the carbon surface~\citep{Singh10,Singh11b,Rajput12}. As a consequence, increasing the ion density in the center of the pores compared to that of the first layer close to the carbon might explain the global increase in diffusion coefficient.    
 
\begin{figure}[ht!]
\includegraphics[scale=0.26]{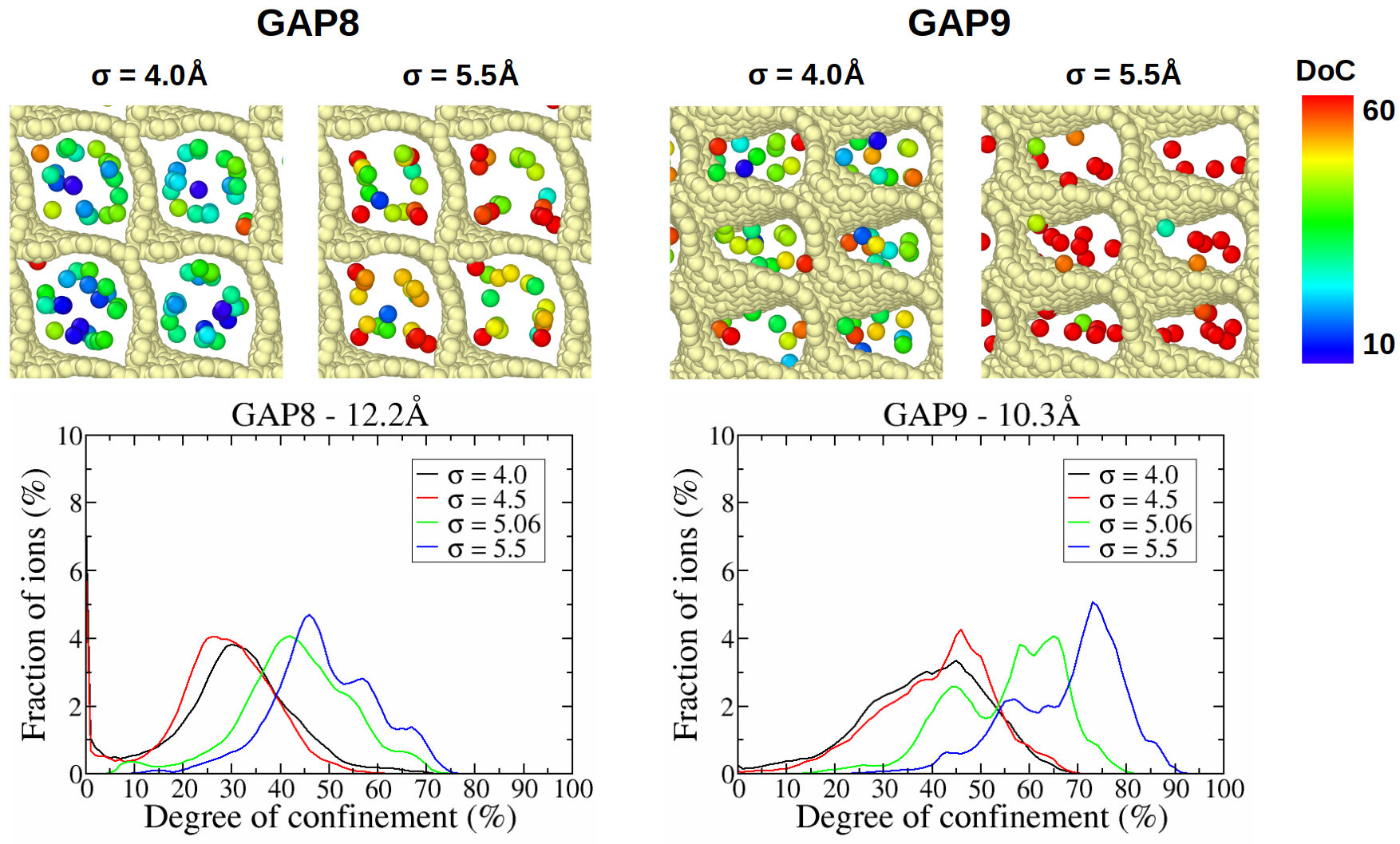}
\caption{Top: Snapshots of anions located in typical adsorption sites in the GAP8 and GAP9 structures. These snapshots were generated using the OVITO software.~\citep{OVITO} Only carbon atoms (in light yellow) and anions (colored according to their instantaneous DoC) are shown for clarity. Bottom: Distribution of the DoC experienced by anions confined in these porous carbons.}
\label{fig:DoC-GAPs-8-9}
\end{figure}

\begin{figure}[ht!]
\includegraphics[scale=0.26]{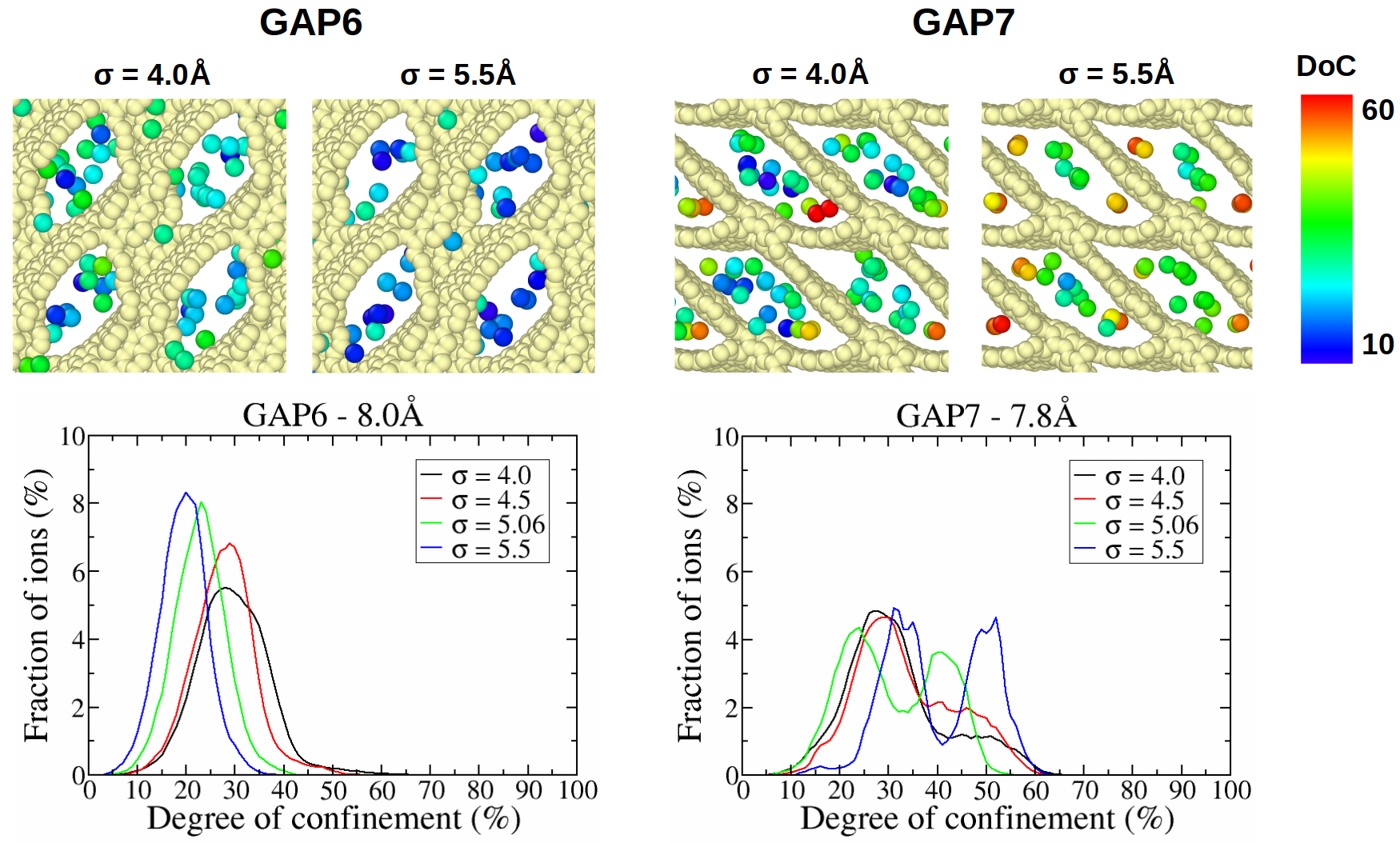}
\caption{Top: Snapshots of anions located in typical adsorption sites in the GAP8 and GAP9 structures. These snapshots were generated using the OVITO software.~\citep{OVITO} Only carbon atoms (in light yellow) and anions (colored according to their instantaneous DoC) are shown for clarity. Bottom: Distribution of the DoC experienced by anions confined in these porous carbons.}
\label{fig:DoC-GAPs-6-7}
\end{figure}

To generalise this analysis to carbon structures where the main directions of diffusion or packing are not $x$, $y$ or $z$, we calculate degrees of confinement (DoCs) of the anions and see how these quantities change with $\sigma_{\rm Anion}$. The degree of confinement, as defined by Merlet~\emph{et~al.}~\citep{Merlet13d}, is the percentage of the solid angle around the ion which is occupied by the carbon atoms, normalized by the maximal value taken by this quantity. As such, the DoC depends both on the number of carbons surrounding an ion and on each ion-carbon distance. Figures~\ref{fig:DoC-GAPs-8-9} and \ref{fig:DoC-GAPs-6-7} show distributions of the DoC experienced by anions confined in a few selected carbons. Distributions for the remaining carbons are given in figure~S5. For almost all the carbons, with the exception of GAP6, the DoCs increase when $\sigma_{\rm Anion}$ increases. At first glance, this is surprising as we would except that larger anions would not be able to enter some of the pores or occupy sites with less confinement. To interpret this result and understand the different behavior of GAP6, we look more closely at some of the carbons, namely GAP6, GAP7, GAP8 and GAP9. 

Figure~\ref{fig:DoC-GAPs-8-9} shows the distributions of the DoC experienced by anions confined in GAP8 and GAP9 carbons as well as snapshots of the anions in typical adsorption sites. Anions are colored according to their instantaneous DoC. GAP8 is the carbon with the largest pore size (12.2~\r{A}). The distribution of DoCs show a large peak around 25-30\% for small $\sigma_{\rm Anion}$  and a main peak with shoulders at very high confinement for large $\sigma_{\rm Anion}$. From the snapshots we can see that the lowest DoCs correspond to anions located closer to the center of the pores while the largest DoCs observed in the case of large anions result from the fact that ions are pushed closer to the pore walls and in the corners. The anions in the corners probably diffuse slower than the ones close to the center of the pores which explains the increase of diffusion coefficients for smaller anions. GAP9 is a carbon with a smaller pore size (10.3~\r{A}) having a much less regular shape. With this carbon we see a change from a wide peak for small anions to a bi-modal distribution for large anions. In the case of the small $\sigma_{\rm Anion}$, it seems that the ions occupy more of the pore space compared to the case of the larger anions where the ions seem to be located in a more ordered fashion. A larger mobility of the small ions in the direction perpendicular to the carbon surface could explain why the distribution is wider in this case compared to the curves for larger ions.

Figure~\ref{fig:DoC-GAPs-6-7} shows the distributions of the DoC experienced by anions confined in GAP6 and GAP7 carbons as well as snapshots of the anions in typical adsorption sites. These carbons are the ones with the smallest pore sizes (7.8 and 8.0~\r{A} for GAP6 and GAP7 respectively). While these carbons have similar pore sizes they have very different topologies, GAP6 has a relatively rectangular shape while GAP7 has diamond-shaped pores. In GAP6, the pore surface is relatively smooth and there are no adsorption sites with especially high DoC. As a consequence, for all anion sizes and pore populations, the ions occupy similar sites. The peak in the distribution of DoCs shifts to lower confinement as the anion size increases, this is probably simply due to an increase in the anion-carbon distance as $\sigma_{\rm Anion}$ increases. It is important to note that while the diffusion coefficient still increases with the ion population, the slope of this increase is much less pronounced than for most of the carbons. The case of GAP7 is very different with the existence of two well identified adsorption sites: the anions can sit in the center of the pore or very confined in the small corners. The distributions of DoCs always show two clear peaks but the relative populations of more confined and less confined ions are varying with the anion size. 

Overall, it seems that the behavior of the diffusion coefficient of the anions is well correlated with the variations in the DoCs. It is difficult to go beyond as the variation of the local diffusion coefficient with the DoC is not known and probably not linear. The existence of exchange between the various positions in the pores also makes it very difficult to analyse the anions trajectories in more depth. This analysis also showed how important the shape of the pores is. Indeed, carbons with very similar average pore sizes, for example GAP6 and GAP7, can show very different behaviors.

\subsection{Effect of the pore size on the confined properties}

After investigating the effect of the anion size on the local structure and dynamics through the DoCs and diffusion coefficients, we now turn to an exploration of the impact of the pore size on the same properties. To this aim we apply a scaling factor of 1.2 and 1.4 on two carbon structures with the same average pore size, namely GAP8 and QMD4x. We keep $\sigma_{\rm Anion}$ constant and equal to 5.06~\r{A}. The initial pore size distributions and their evolution with the scaling procedure are shown in Figure~S6. It is important to note that while this method allows us to focus on a change of pore size at constant topology, this generates structures with different densities than the original ones and larger C-C bonds. 

\begin{figure}[ht!]
\includegraphics[scale=0.26]{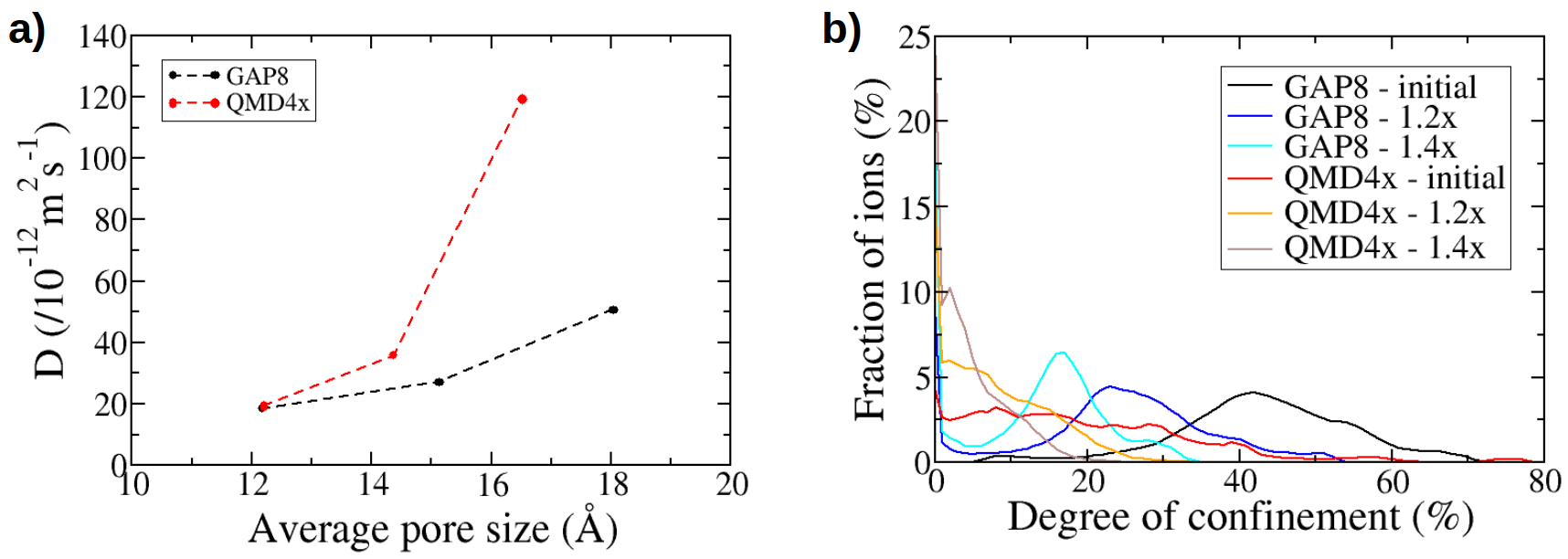}
\caption{a) Diffusion coefficients for anions confined in scaled carbons with different pore sizes, and b) Distribution of the DoC experienced by anions confined in these porous carbons.}
\label{fig:Diff-DoC-scaling}
\end{figure}

Figure~\ref{fig:Diff-DoC-scaling} shows the diffusion coefficients as a function of pore size and the distribution of DoCs for carbons GAP8 and QMD4x and their scaled counterparts. We observe a very clear increase of the diffusion coefficient with the pore size as expected. We note that the diffusion coefficients for the ordered GAP8 carbon are always lower than those of the disordered QMD4x carbon. For the GAP8 carbon, the distributions of DoCs show a single peak towards lower confinement when the pore size increases. Compared to the initial GAP carbons, we can see much more environments at DoCs very close to zero in the scaled carbons. For the QMD4x carbon, the distributions of DoCs are much wider with no clear peak but the shift to lower confinements is also observed with the pore size increase. The absence of a clear peak is probably the result of a lack of clear defined geometries in this highly disordered carbon. It is worth pointing out that the DoCs are usually lower for the QMD4x carbon with respect to the GAP8 structure, in agreement with the larger diffusion coefficients calculated in the disordered carbon.

The results provided by the simulations with scaled carbon structures are concordant with the observations made for the anion size effect.

\subsection{Comparison of topologically different carbon structures}

After discussing the effects of the anion size and the pore size on the electrolyte properties under confinement, we now turn to an attempt to rationalize the correlations between the porous structure and the properties of the confined electrolytes. There are a number of properties used to characterise porous structures, none being ideal as it is very challenging to describe a complex structure with only a few key parameters. Here, we use Poreblazer~\citep{Sarkisov11} to extract the properties of the carbon structures. We determined the Pore Size Distributions (PSD) and use them to calculate the average pore sizes (see Figures~\ref{fig:carbons} and~S7). While not being sufficient, the average pore size is very often used to compare carbons and was shown to correlate with some electrochemical properties~\cite{Chmiola06,Raymundo-Pinero06}. The other properties we are going to use are the Pore Limiting Diameter, i.e.~the smallest opening along the pore that a molecule needs to cross in order to diffuse through this material, and the Maximum Pore Diameter, i.e.~the largest opening along the pore. The position of these two quantities with respect to the PSD are shown on two carbon structures in Supplementary Information. In this part, we analyse molecular simulations done with the initial coarse-grained model for the ionic liquid ($\sigma_{\rm Anion}$~=~5.06~\r{A}) in contact with 14 carbon structures (10 GAPs and 4 QMDs).

\begin{figure}[ht!]
\includegraphics[scale=0.26]{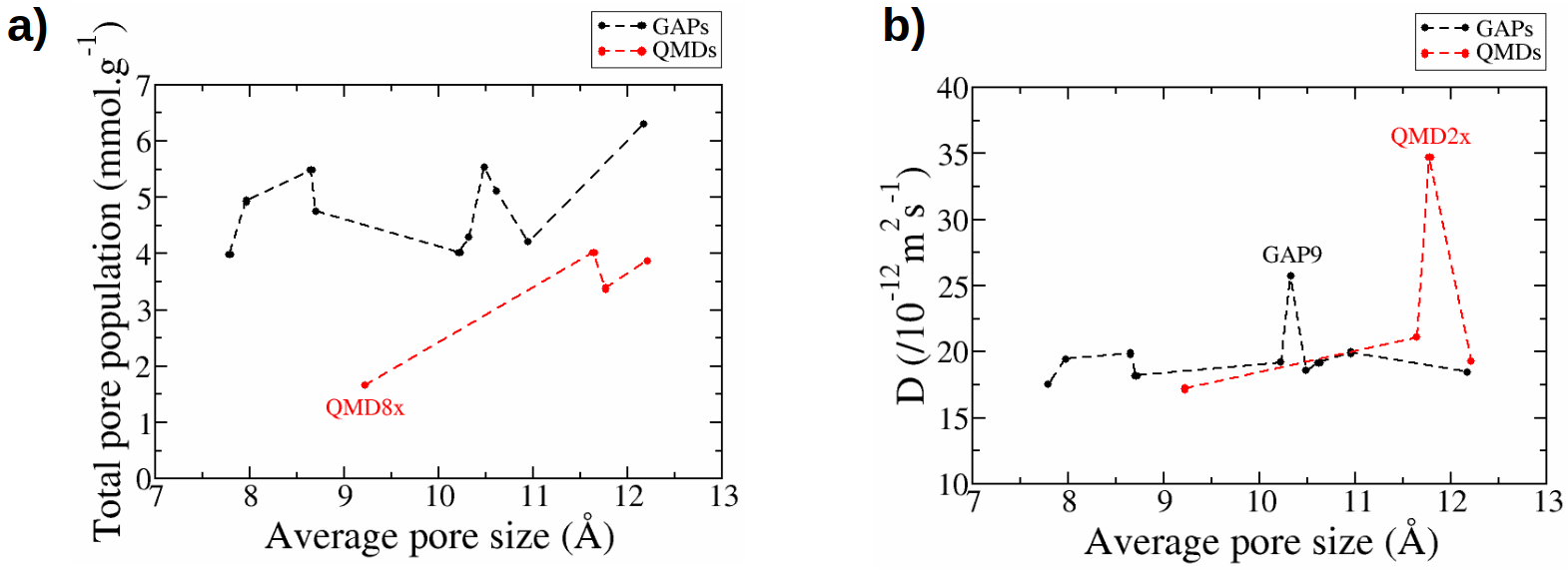}
\caption{a) Total pore populations, and b) diffusion coefficients as a function of average pore size for ions confined in the various carbons.}
\label{fig:TPP-Diff-poresize}
\end{figure}

\begin{figure}[ht!]
\includegraphics[scale=0.28]{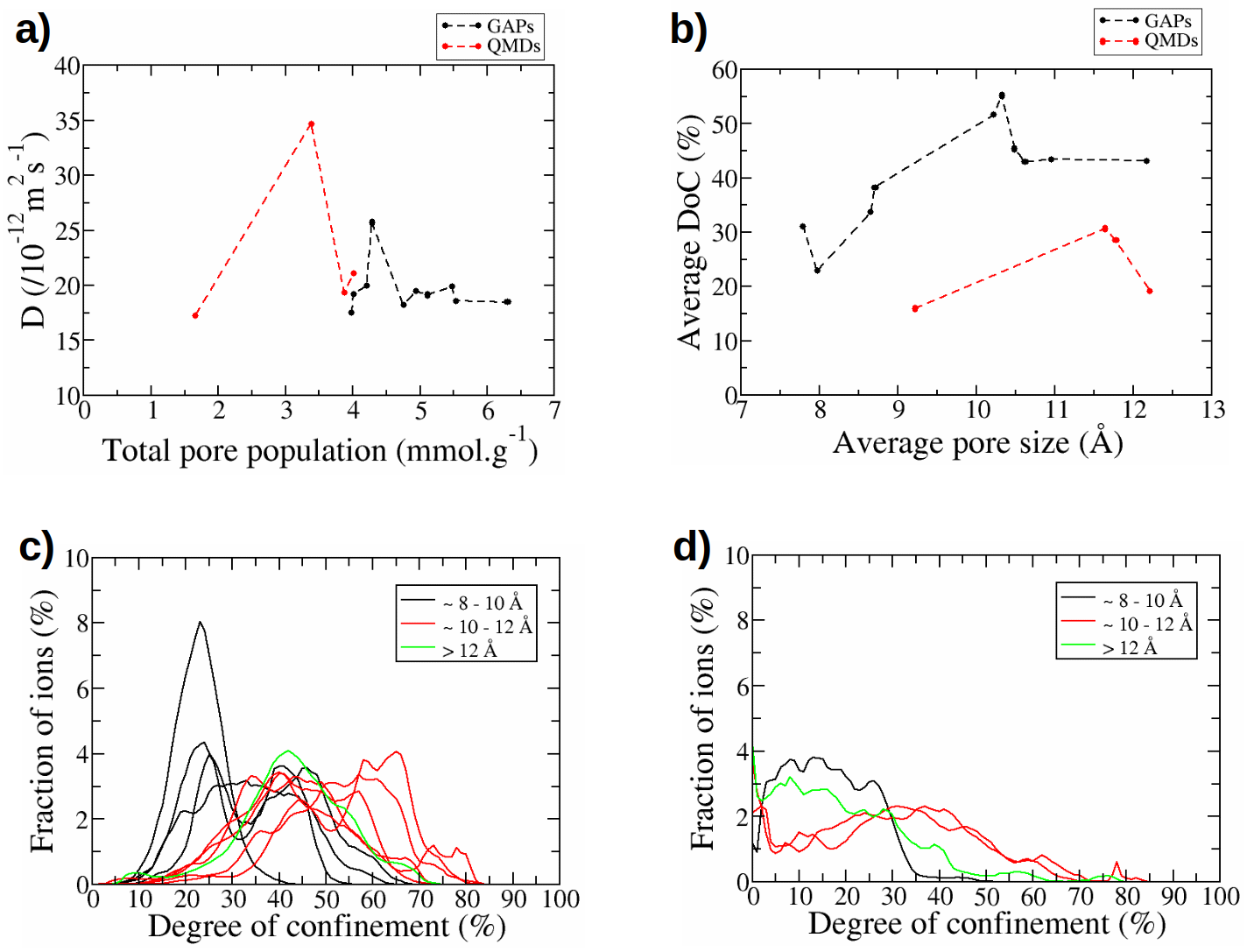}
\caption{a) Diffusion coefficients as a function of the total pore populations, b) degree of confinement as a function of average pore size, c) d) distributions of DoCs for the anions confined in all the nanoporous carbons studied here.}
\label{fig:Diff-DoC-poresize}
\end{figure}

Figure~\ref{fig:TPP-Diff-poresize} shows the total pore population and the anion diffusion coefficient for all the carbons studied here. The first thing we observe is that the total pore population is always lower for QMDs compared to GAPs. The QMDs are much more disordered which might prevent an optimal packing of the ions in the porosity. We note that one of the disordered carbons, QMD8x, contains much less ions (almost three times less) than the others. This carbon is the only QMD carbon that does not contain a big pore (its PSD does not show any maximum close to 14~\r{A}). Another interesting point is that for the GAP carbons which span a larger range of pore sizes, no monotonous increase of the total pore population with the pore size is observed but some structures seem to allow for larger quantities of ions to pack in. We also note that going across carbons with different topologies and going from small to large pore sizes, the diffusion coefficient is not monotonous, as already observed by Wang~\emph{et al.}~\citep{ChenluWang} in slit pores, and not correlated with the total pore population, as is clear from Figure~\ref{fig:Diff-DoC-poresize}a. This underlines the importance of pore topology and the inability of the average pore size as a single parameter to characterize the porous structure. Interestingly the highest diffusion coefficients are observed for the GAP9 and QMD2x carbons, the two carbons which show the largest diffusion along $x$ and $y$ directions, i.e. perpendicularly to the main direction (see Figure~S8 for the different components of the diffusion coefficients). This suggests that diffusion in three dimensions and not only in a channel is important for a better diffusion.  

We now calculate the DoCs in the different carbons to try to get insights into these variations. Figure~\ref{fig:Diff-DoC-poresize} shows the distributions of DoCs for anions adsorbed in all the carbons (Figure~\ref{fig:Diff-DoC-poresize}c)  as well as the variation of the DoC with the pore size (Figure~\ref{fig:Diff-DoC-poresize}d). One might expect that the DoC increases when the pore size decreases but this is not the case. Actually, the confinement seems to be lower for very small pore sizes, increases for intermediate pore sizes and decreases again above 12~\r{A}. This trend is seen clearly in Figures~\ref{fig:Diff-DoC-poresize}c and~\ref{fig:Diff-DoC-poresize}d where the curves for different carbons are colored according to the range of pore sizes they belong to. In particular, the highest DoCs are observed for intermediate average pore sizes between 10~\r{A} and 12~\r{A}. Interestingly, the same trend is seen for GAPs and QMDs even if the DoC distributions for QMDs are much wider.  

Looking at the ordered GAP carbons and thinking about geometrical descriptors for the pore topologies, one idea that comes to mind is to use the ratio of the longest distance over the shortest distance defining the main pore. If the pore is square-like then this ratio will be close to one. If the pore is more diamond-shaped then this ratio will be larger than one. The more distorted the pore is, the larger the ratio will become. The value of the ratio is thus very dependent on the pore shape and we call it ``form factor" in the remainder of this article. These shortest and longest distances are not always well defined, especially for disordered carbons, so, as a proxy, we will use the maximum pore diameter and the pore limiting diameter determined using Poreblazer~\citep{Sarkisov11}. 

\begin{figure}[ht!]
\includegraphics[scale=0.26]{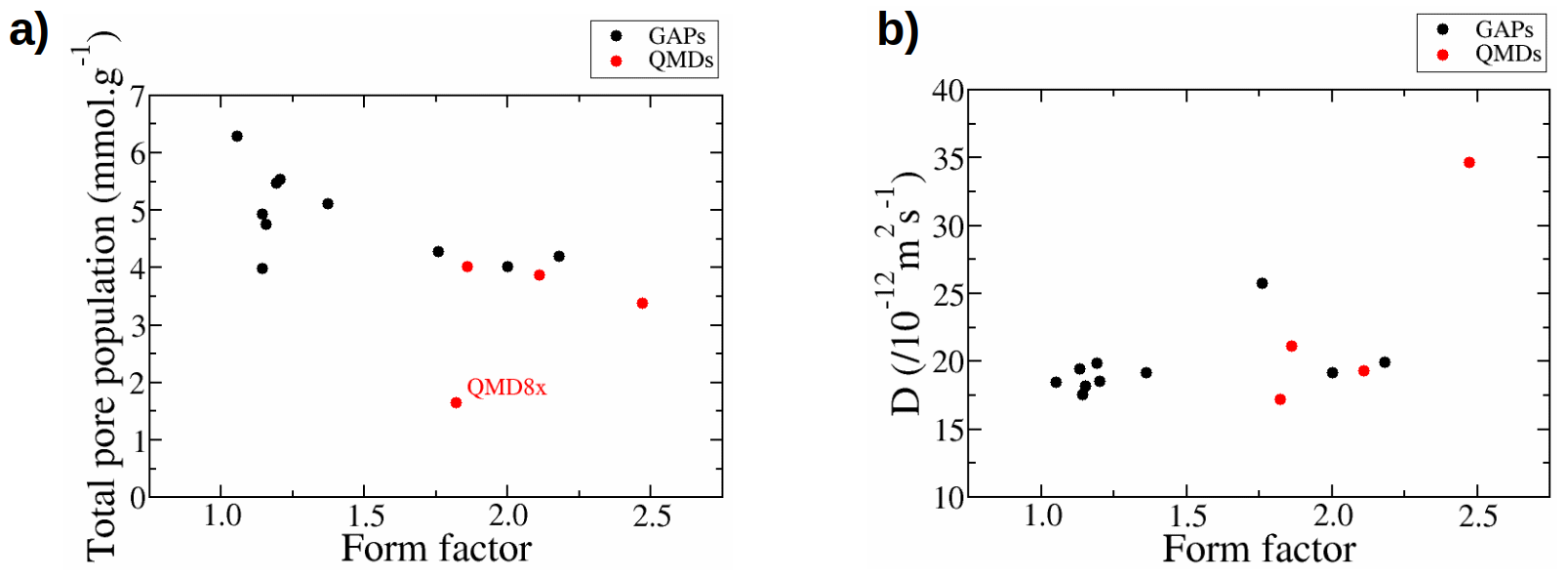}
\caption{a) Total pore populations, and b) diffusion coefficients as a function of the form factor defined as the ratio between the maximum pore diameter and the pore limiting diameter.}
\label{fig:TPP-Diff-MPDoverPLD}
\end{figure}

Plots showing the total pore population and the diffusion coefficients as a function of the form factor are given in Figure~\ref{fig:TPP-Diff-MPDoverPLD}. From this figure, it seems that this descriptor is indeed correlated with the total pore population for a range of carbons, including three of the disordered carbons. QMD8x is again an outlier showing a very low quantity of adsorbed ions. Close to the value of 1.0 for the form factor, the dispersion is larger. In this region, the effect of the actual pore size is probably more important as the pore is more square or sphere like. On the contrary, no clear correlation can be established between the diffusion coefficients and the form factor. It is worth noting though that the actual values of the diffusion coefficients are relatively similar with many values around 20~10$^{-12}$~m$^2$~s$^{-1}$ which makes it more challenging to identify trends. 

Overall, this comparison between carbons having different topologies have underlined the fact that smaller pores do not necessarily lead to larger confinement and suggested that more relevant geometrical descriptors than the pore sizes could be used. A further step, out of the scope of the present work and requiring a larger database, would be to characterize the topology in a more complete way, for example using a pore recognition approach~\citep{Lee17}.

\section{Conclusion}

We have carried out a methodical molecular dynamics simulations study of ionic liquid adsorption into nanoporous carbons to investigate the correlation between ion size, pore size, porous structure and structural and dynamical properties. We have shown that a change in anion size does not affect the diffusion coefficients in the same way in the bulk or under confinement. Moreover, under confinement, we have shown that for a given carbon structure, the diffusion coefficient increases when the total pore population increases. This surprising result was explained by looking at the position of the ions within the pores. Assuming that more confined ions diffuse slower, as proposed in the literature, the increase in ion population leads to more ions being less confined, located in a more central position in the pores, and results in a diffusion coefficient increase. The degree of confinement of the ions was also analysed in scaled carbon structures which allowed us to explore a change of pore size while keeping a constant topology. The results are concordant with the ones obtained for varying anion sizes. We then tried to apply our analysis to the study of the same pure ionic liquid in contact with 14 carbon structures, ordered and disordered. This task appeared to be more challenging but we could highlight some interesting features. It was shown that the degree of confinement does not necessarily increase with the decrease in the average pore size. Actually, it seems that the pore sizes could be divided in three sets: the small pore sizes below 10~\r{A}, the intermediate pore sizes between 10~\r{A} and 12~\r{A} which correspond to the largest confinements observed and the largest pore sizes above 12~\r{A}. In addition, a correlation between the total pore population and the ratio between the maximum pore diameter and the pore limiting diameter was observed. In the future, the characterization of the porous structures would benefit a lot from more complete approaches, for example using pore recognition.     
\section*{Data availability}

The data corresponding to the plots reported in this paper, as well as an example input file for LAMMPS, are available in the Zenodo repository with identifier 10.5281/zenodo.3407266.

\begin{acknowledgement}

This project has received funding from the European Research Council (ERC) under the European Union’s Horizon 2020 research and innovation programme (grant agreement no. 714581). This work was granted access to the HPC resources of CALMIP supercomputing center under the allocation P17037. The authors acknowledge Pierre-Louis Taberna, Michelle Liu, Berend Smit, Mathieu Salanne, Benjamin Rotenberg, and Nidhal Ganfoud for useful discussions.

\end{acknowledgement}

\begin{suppinfo}
Lennard-Jones potentials, snapshots of the carbons studied, additional diffusion coefficients and DoC plots, pore size distributions of the scaled carbons, average pore sizes, and $x$/$y$/$z$ components of the diffusion coefficients are given in Supplementary Information.
\end{suppinfo}


\begin{mcitethebibliography}{40}
\providecommand*\natexlab[1]{#1}
\providecommand*\mciteSetBstSublistMode[1]{}
\providecommand*\mciteSetBstMaxWidthForm[2]{}
\providecommand*\mciteBstWouldAddEndPuncttrue
  {\def\EndOfBibitem{\unskip.}}
\providecommand*\mciteBstWouldAddEndPunctfalse
  {\let\EndOfBibitem\relax}
\providecommand*\mciteSetBstMidEndSepPunct[3]{}
\providecommand*\mciteSetBstSublistLabelBeginEnd[3]{}
\providecommand*\EndOfBibitem{}
\mciteSetBstSublistMode{f}
\mciteSetBstMaxWidthForm{subitem}{(\alph{mcitesubitemcount})}
\mciteSetBstSublistLabelBeginEnd
  {\mcitemaxwidthsubitemform\space}
  {\relax}
  {\relax}

\bibitem[Zhong \latin{et~al.}(2015)Zhong, Deng, Hu, Qiao, Zhang, and
  Zhang]{Zhong15}
Zhong,~C.; Deng,~Y.; Hu,~W.; Qiao,~J.; Zhang,~L.; Zhang,~J. A review of
  electrolyte materials and compositions for electrochemical supercapacitors.
  \emph{Chem. Soc. Rev.} \textbf{2015}, \emph{44}, 7484--7539\relax
\mciteBstWouldAddEndPuncttrue
\mciteSetBstMidEndSepPunct{\mcitedefaultmidpunct}
{\mcitedefaultendpunct}{\mcitedefaultseppunct}\relax
\EndOfBibitem
\bibitem[B\'eguin \latin{et~al.}(2014)B\'eguin, Presser, Balducci, and
  Frackowiak]{Beguin14}
B\'eguin,~F.; Presser,~V.; Balducci,~A.; Frackowiak,~E. Carbons and
  Electrolytes for Advanced Supercapacitors. \emph{Adv. Mater.} \textbf{2014},
  \emph{26}, 2219--2251\relax
\mciteBstWouldAddEndPuncttrue
\mciteSetBstMidEndSepPunct{\mcitedefaultmidpunct}
{\mcitedefaultendpunct}{\mcitedefaultseppunct}\relax
\EndOfBibitem
\bibitem[Brandt \latin{et~al.}(2013)Brandt, Pohlmann, Varzi, Balducci, and
  Passerini]{Brandt13}
Brandt,~A.; Pohlmann,~S.; Varzi,~A.; Balducci,~A.; Passerini,~S. Ionic liquids
  in supercapacitors. \emph{MRS Bull.} \textbf{2013}, \emph{38}, 554--559\relax
\mciteBstWouldAddEndPuncttrue
\mciteSetBstMidEndSepPunct{\mcitedefaultmidpunct}
{\mcitedefaultendpunct}{\mcitedefaultseppunct}\relax
\EndOfBibitem
\bibitem[Lewandowski \latin{et~al.}(2010)Lewandowski, Olejniczak, Galinski, and
  Stepniak]{Lewandowski10}
Lewandowski,~A.; Olejniczak,~A.; Galinski,~M.; Stepniak,~I. Performance of
  carbon-carbon supercapacitors based on organic, aqueous and ionic liquid
  electrolytes. \emph{J. Power Sources} \textbf{2010}, \emph{195}, 5814 --
  5819\relax
\mciteBstWouldAddEndPuncttrue
\mciteSetBstMidEndSepPunct{\mcitedefaultmidpunct}
{\mcitedefaultendpunct}{\mcitedefaultseppunct}\relax
\EndOfBibitem
\bibitem[Simon and Gogotsi(2013)Simon, and Gogotsi]{Simon13}
Simon,~P.; Gogotsi,~Y. Capacitive Energy Storage in Nanostructured
  Carbon--Electrolyte Systems. \emph{Accounts Chem. Res.} \textbf{2013},
  \emph{46}, 1094--1103\relax
\mciteBstWouldAddEndPuncttrue
\mciteSetBstMidEndSepPunct{\mcitedefaultmidpunct}
{\mcitedefaultendpunct}{\mcitedefaultseppunct}\relax
\EndOfBibitem
\bibitem[Liu \latin{et~al.}(2017)Liu, Zhang, Song, and Li]{Liu17}
Liu,~T.; Zhang,~F.; Song,~Y.; Li,~Y. Revitalizing carbon supercapacitor
  electrodes with hierarchical porous structures. \emph{J. Mater. Chem. A}
  \textbf{2017}, \emph{5}, 17705--17733\relax
\mciteBstWouldAddEndPuncttrue
\mciteSetBstMidEndSepPunct{\mcitedefaultmidpunct}
{\mcitedefaultendpunct}{\mcitedefaultseppunct}\relax
\EndOfBibitem
\bibitem[Fic \latin{et~al.}(2018)Fic, Platek, Piwek, and Frackowiak]{Fic18}
Fic,~K.; Platek,~A.; Piwek,~J.; Frackowiak,~E. Sustainable materials for
  electrochemical capacitors. \emph{Materials Today} \textbf{2018}, \emph{21},
  437 -- 454\relax
\mciteBstWouldAddEndPuncttrue
\mciteSetBstMidEndSepPunct{\mcitedefaultmidpunct}
{\mcitedefaultendpunct}{\mcitedefaultseppunct}\relax
\EndOfBibitem
\bibitem[Lee \latin{et~al.}(2017)Lee, Perez-Martinez, Smith, and
  Perkin]{Lee17b}
Lee,~A.~A.; Perez-Martinez,~C.~S.; Smith,~A.~M.; Perkin,~S. Underscreening in
  concentrated electrolytes. \emph{Faraday Discuss.} \textbf{2017}, \emph{199},
  239--259\relax
\mciteBstWouldAddEndPuncttrue
\mciteSetBstMidEndSepPunct{\mcitedefaultmidpunct}
{\mcitedefaultendpunct}{\mcitedefaultseppunct}\relax
\EndOfBibitem
\bibitem[Lhermerout and Perkin(2018)Lhermerout, and Perkin]{Lhermerout18}
Lhermerout,~R.; Perkin,~S. Nanoconfined ionic liquids: Disentangling
  electrostatic and viscous forces. \emph{Phys. Rev. Fluids} \textbf{2018},
  \emph{3}, 014201\relax
\mciteBstWouldAddEndPuncttrue
\mciteSetBstMidEndSepPunct{\mcitedefaultmidpunct}
{\mcitedefaultendpunct}{\mcitedefaultseppunct}\relax
\EndOfBibitem
\bibitem[Tsai \latin{et~al.}(2014)Tsai, Taberna, and Simon]{Tsai14}
Tsai,~W.-Y.; Taberna,~P.-L.; Simon,~P. {E}lectrochemical {Q}uartz {C}rystal
  {M}icrobalance ({EQCM}) Study of Ion Dynamics in Nanoporous Carbons. \emph{J.
  Am. Chem. Soc.} \textbf{2014}, \emph{136}, 8722--8728\relax
\mciteBstWouldAddEndPuncttrue
\mciteSetBstMidEndSepPunct{\mcitedefaultmidpunct}
{\mcitedefaultendpunct}{\mcitedefaultseppunct}\relax
\EndOfBibitem
\bibitem[Forse \latin{et~al.}(2016)Forse, Merlet, Griffin, and Grey]{Forse16}
Forse,~A.~C.; Merlet,~C.; Griffin,~J.~M.; Grey,~C.~P. New Perspectives on the
  Charging Mechanisms of Supercapacitors. \emph{J. Am. Chem. Soc.}
  \textbf{2016}, \emph{138}, 5731--5744\relax
\mciteBstWouldAddEndPuncttrue
\mciteSetBstMidEndSepPunct{\mcitedefaultmidpunct}
{\mcitedefaultendpunct}{\mcitedefaultseppunct}\relax
\EndOfBibitem
\bibitem[Forse \latin{et~al.}(2015)Forse, Griffin, Merlet, Bayley, Wang, Simon,
  and Grey]{Forse15}
Forse,~A.~C.; Griffin,~J.~M.; Merlet,~C.; Bayley,~P.~M.; Wang,~H.; Simon,~P.;
  Grey,~C.~P. {NMR} Study of Ion Dynamics and Charge Storage in Ionic Liquid
  Supercapacitors. \emph{J. Am. Chem. Soc.} \textbf{2015}, \emph{137},
  7231--7242\relax
\mciteBstWouldAddEndPuncttrue
\mciteSetBstMidEndSepPunct{\mcitedefaultmidpunct}
{\mcitedefaultendpunct}{\mcitedefaultseppunct}\relax
\EndOfBibitem
\bibitem[Forse \latin{et~al.}(2017)Forse, Griffin, Merlet, Carreteo-Gonzalez,
  Raji, Trease, and Grey]{Forse17}
Forse,~A.~C.; Griffin,~J.~M.; Merlet,~C.; Carreteo-Gonzalez,~J.;
  Raji,~A.-R.~O.; Trease,~N.~M.; Grey,~C.~P. Direct observation of ion dynamics
  in supercapacitor electrodes using in situ diffusion NMR spectroscopy.
  \emph{Nature Ener.} \textbf{2017}, \emph{2}, 16216\relax
\mciteBstWouldAddEndPuncttrue
\mciteSetBstMidEndSepPunct{\mcitedefaultmidpunct}
{\mcitedefaultendpunct}{\mcitedefaultseppunct}\relax
\EndOfBibitem
\bibitem[Kornyshev(2007)]{Kornyshev07}
Kornyshev,~A.~A. Double-Layer in Ionic Liquids:  Paradigm Change? \emph{J.
  Phys. Chem. B} \textbf{2007}, \emph{111}, 5545--5557\relax
\mciteBstWouldAddEndPuncttrue
\mciteSetBstMidEndSepPunct{\mcitedefaultmidpunct}
{\mcitedefaultendpunct}{\mcitedefaultseppunct}\relax
\EndOfBibitem
\bibitem[Huang \latin{et~al.}(2008)Huang, Sumpter, and Meunier]{Huang08}
Huang,~J.; Sumpter,~B.~G.; Meunier,~V. Theoretical Model for Nanoporous Carbon
  Supercapacitors. \emph{Angew. Chem. Int. Edit.} \textbf{2008}, \emph{47},
  520--524\relax
\mciteBstWouldAddEndPuncttrue
\mciteSetBstMidEndSepPunct{\mcitedefaultmidpunct}
{\mcitedefaultendpunct}{\mcitedefaultseppunct}\relax
\EndOfBibitem
\bibitem[Kondrat and Kornyshev(2011)Kondrat, and Kornyshev]{Kondrat11}
Kondrat,~S.; Kornyshev,~A. Superionic state in double-layer capacitors with
  nanoporous electrodes. \emph{J. Phys.: Condens. Matter} \textbf{2011},
  \emph{23}, 022201\relax
\mciteBstWouldAddEndPuncttrue
\mciteSetBstMidEndSepPunct{\mcitedefaultmidpunct}
{\mcitedefaultendpunct}{\mcitedefaultseppunct}\relax
\EndOfBibitem
\bibitem[Merlet \latin{et~al.}(2013)Merlet, Rotenberg, Madden, and
  Salanne]{Merlet13c}
Merlet,~C.; Rotenberg,~B.; Madden,~P.~A.; Salanne,~M. Computer simulations of
  ionic liquids at electrochemical interfaces. \emph{Phys. Chem. Chem. Phys.}
  \textbf{2013}, \emph{15}, 15781--15792\relax
\mciteBstWouldAddEndPuncttrue
\mciteSetBstMidEndSepPunct{\mcitedefaultmidpunct}
{\mcitedefaultendpunct}{\mcitedefaultseppunct}\relax
\EndOfBibitem
\bibitem[Fedorov and Kornyshev(2014)Fedorov, and Kornyshev]{Fedorov14}
Fedorov,~M.~V.; Kornyshev,~A.~A. Ionic Liquids at Electrified Interfaces.
  \emph{Chem. Rev.} \textbf{2014}, \emph{114}, 2978--3036\relax
\mciteBstWouldAddEndPuncttrue
\mciteSetBstMidEndSepPunct{\mcitedefaultmidpunct}
{\mcitedefaultendpunct}{\mcitedefaultseppunct}\relax
\EndOfBibitem
\bibitem[Burt \latin{et~al.}(2016)Burt, Breitsprecher, Daffos, Taberna, Simon,
  Birkett, Zhao, Holm, and Salanne]{Burt16}
Burt,~R.; Breitsprecher,~K.; Daffos,~B.; Taberna,~P.-L.; Simon,~P.;
  Birkett,~G.; Zhao,~X.~S.; Holm,~C.; Salanne,~M. Capacitance of Nanoporous
  Carbon-Based Supercapacitors Is a Trade-Off between the Concentration and the
  Separability of the Ions. \emph{J. Phys. Chem. Lett.} \textbf{2016},
  \emph{7}, 4015--4021\relax
\mciteBstWouldAddEndPuncttrue
\mciteSetBstMidEndSepPunct{\mcitedefaultmidpunct}
{\mcitedefaultendpunct}{\mcitedefaultseppunct}\relax
\EndOfBibitem
\bibitem[He \latin{et~al.}(2016)He, Qiao, Vatamanu, Borodin, Bedrov, Huang, and
  Sumpter]{He16}
He,~Y.; Qiao,~R.; Vatamanu,~J.; Borodin,~O.; Bedrov,~D.; Huang,~J.;
  Sumpter,~B.~G. The Importance of Ion Packing on the Dynamics of Ionic Liquids
  during Micropore Charging. \emph{J. Phys. Chem. Lett.} \textbf{2016},
  \emph{7}, 36--42\relax
\mciteBstWouldAddEndPuncttrue
\mciteSetBstMidEndSepPunct{\mcitedefaultmidpunct}
{\mcitedefaultendpunct}{\mcitedefaultseppunct}\relax
\EndOfBibitem
\bibitem[Roy and Maroncelli(2010)Roy, and Maroncelli]{Roy10b}
Roy,~D.; Maroncelli,~M. An Improved Four-Site Ionic Liquid Model. \emph{J.
  Phys. Chem. B} \textbf{2010}, \emph{114}, 12629--12631\relax
\mciteBstWouldAddEndPuncttrue
\mciteSetBstMidEndSepPunct{\mcitedefaultmidpunct}
{\mcitedefaultendpunct}{\mcitedefaultseppunct}\relax
\EndOfBibitem
\bibitem[Cole and Klein(1983)Cole, and Klein]{Cole83}
Cole,~M.~W.; Klein,~J.~R. The interaction between noble gases and the basal
  plane surface of graphite. \emph{Surf. Sci.} \textbf{1983}, \emph{124}, 547
  -- 554\relax
\mciteBstWouldAddEndPuncttrue
\mciteSetBstMidEndSepPunct{\mcitedefaultmidpunct}
{\mcitedefaultendpunct}{\mcitedefaultseppunct}\relax
\EndOfBibitem
\bibitem[Humphrey \latin{et~al.}(1996)Humphrey, Dalke, and Schulten]{VMD}
Humphrey,~W.; Dalke,~A.; Schulten,~K. VMD-VisualMolecularDynamics. \emph{J.
  Mol. Graphics} \textbf{1996}, \emph{14}, 33--38\relax
\mciteBstWouldAddEndPuncttrue
\mciteSetBstMidEndSepPunct{\mcitedefaultmidpunct}
{\mcitedefaultendpunct}{\mcitedefaultseppunct}\relax
\EndOfBibitem
\bibitem[Deringer \latin{et~al.}(2018)Deringer, Merlet, Hu, Lee, Kattirtzi,
  Pecher, Cs\'anyi, Elliott, and Grey]{Deringer18}
Deringer,~V.~L.; Merlet,~C.; Hu,~Y.; Lee,~T.~H.; Kattirtzi,~J.~A.; Pecher,~O.;
  Cs\'anyi,~G.; Elliott,~S.~R.; Grey,~C.~P. Towards an atomistic understanding
  of disordered carbon electrode materials. \emph{Chem. Commun.} \textbf{2018},
  \emph{54}, 5988--5991\relax
\mciteBstWouldAddEndPuncttrue
\mciteSetBstMidEndSepPunct{\mcitedefaultmidpunct}
{\mcitedefaultendpunct}{\mcitedefaultseppunct}\relax
\EndOfBibitem
\bibitem[Deringer and Cs\'anyi(2017)Deringer, and Cs\'anyi]{Deringer17}
Deringer,~V.~L.; Cs\'anyi,~G. Machine learning based interatomic potential for
  amorphous carbon. \emph{Phys. Rev. B} \textbf{2017}, \emph{95}, 094203\relax
\mciteBstWouldAddEndPuncttrue
\mciteSetBstMidEndSepPunct{\mcitedefaultmidpunct}
{\mcitedefaultendpunct}{\mcitedefaultseppunct}\relax
\EndOfBibitem
\bibitem[Palmer \latin{et~al.}(2010)Palmer, Llobet, Yeon, Fischer, Shi,
  Gogotsi, and Gubbins]{Palmer10}
Palmer,~J.~C.; Llobet,~A.; Yeon,~S.-H.; Fischer,~J.~E.; Shi,~Y.; Gogotsi,~Y.;
  Gubbins,~K.~E. Modeling the structural evolution of carbide-derived carbons
  using quenched molecular dynamics. \emph{Carbon} \textbf{2010}, \emph{48},
  1116 -- 1123\relax
\mciteBstWouldAddEndPuncttrue
\mciteSetBstMidEndSepPunct{\mcitedefaultmidpunct}
{\mcitedefaultendpunct}{\mcitedefaultseppunct}\relax
\EndOfBibitem
\bibitem[Sarkisov and Harrison(2011)Sarkisov, and Harrison]{Sarkisov11}
Sarkisov,~L.; Harrison,~A. Computational structure characterisation tools in
  application to ordered and disordered porous materials. \emph{Mol. Simulat.}
  \textbf{2011}, \emph{37}, 1248--1257\relax
\mciteBstWouldAddEndPuncttrue
\mciteSetBstMidEndSepPunct{\mcitedefaultmidpunct}
{\mcitedefaultendpunct}{\mcitedefaultseppunct}\relax
\EndOfBibitem
\bibitem[Plimpton(1995)]{LAMMPS}
Plimpton,~S. Fast Parallel Algorithms for Short-Range Molecular Dynamics.
  \emph{J. Comp. Phys.} \textbf{1995}, \emph{117}, 1--19\relax
\mciteBstWouldAddEndPuncttrue
\mciteSetBstMidEndSepPunct{\mcitedefaultmidpunct}
{\mcitedefaultendpunct}{\mcitedefaultseppunct}\relax
\EndOfBibitem
\bibitem[Liu \latin{et~al.}(2004)Liu, Harder, and Berne]{Liu04}
Liu,~P.; Harder,~E.; Berne,~B. On the calculation of diffusion coefficients in
  confined fluids and interfaces with an application to the liquid-vapor
  interface of water. \emph{J. Phys. Chem.} \textbf{2004}, \emph{108}, 6595 --
  6602\relax
\mciteBstWouldAddEndPuncttrue
\mciteSetBstMidEndSepPunct{\mcitedefaultmidpunct}
{\mcitedefaultendpunct}{\mcitedefaultseppunct}\relax
\EndOfBibitem
\bibitem[Rotenberg \latin{et~al.}(2007)Rotenberg, Marry, Vuilleumier, Malikova,
  Simon, and Turq]{Rotenberg07}
Rotenberg,~B.; Marry,~V.; Vuilleumier,~R.; Malikova,~N.; Simon,~C.; Turq,~P.
  Water and ions in clays: Unraveling the interlayer/micropore exchange using
  molecular dynamics. \emph{Geochim. et Cosmochim. Acta} \textbf{2007},
  \emph{71}, 5089 -- 5101\relax
\mciteBstWouldAddEndPuncttrue
\mciteSetBstMidEndSepPunct{\mcitedefaultmidpunct}
{\mcitedefaultendpunct}{\mcitedefaultseppunct}\relax
\EndOfBibitem
\bibitem[Rajput \latin{et~al.}(2012)Rajput, Monk, Singh, and Hung]{Rajput12}
Rajput,~N.~N.; Monk,~J.; Singh,~R.; Hung,~F.~R. On the Influence of Pore Size
  and Pore Loading on Structural and Dynamical Heterogeneities of an Ionic
  Liquid Confined in a Slit Nanopore. \emph{J. Phys. Chem. C} \textbf{2012},
  \emph{116}, 5169--5181\relax
\mciteBstWouldAddEndPuncttrue
\mciteSetBstMidEndSepPunct{\mcitedefaultmidpunct}
{\mcitedefaultendpunct}{\mcitedefaultseppunct}\relax
\EndOfBibitem
\bibitem[Wang \latin{et~al.}(2019)Wang, Wang, Lu, He, Huo, Dong, Wei, and
  Zhang]{ChenluWang}
Wang,~C.; Wang,~Y.; Lu,~Y.; He,~H.; Huo,~F.; Dong,~K.; Wei,~N.; Zhang,~S.
  Height-driven structure and thermodynamic properties of confined ionic
  liquids inside carbon nanochannels from molecular dynamics study. \emph{Phys.
  Chem. Chem. Phys.} \textbf{2019}, \emph{21}, 12767--12776\relax
\mciteBstWouldAddEndPuncttrue
\mciteSetBstMidEndSepPunct{\mcitedefaultmidpunct}
{\mcitedefaultendpunct}{\mcitedefaultseppunct}\relax
\EndOfBibitem
\bibitem[Singh \latin{et~al.}(2010)Singh, Monk, and Hung]{Singh10}
Singh,~R.; Monk,~J.; Hung,~F.~R. A Computational Study of the Behavior of the
  Ionic Liquid $[${BMIM}$^+][${PF6}$^-]$ Confined Inside Multiwalled Carbon
  Nanotubes. \emph{J. Phys. Chem. C} \textbf{2010}, \emph{114},
  15478--15485\relax
\mciteBstWouldAddEndPuncttrue
\mciteSetBstMidEndSepPunct{\mcitedefaultmidpunct}
{\mcitedefaultendpunct}{\mcitedefaultseppunct}\relax
\EndOfBibitem
\bibitem[Singh \latin{et~al.}(2011)Singh, Monk, and Hung]{Singh11b}
Singh,~R.; Monk,~J.; Hung,~F.~R. Heterogeneity in the Dynamics of the Ionic
  Liquid $[${BMIM}$^+][${PF}$_6^-]$ Confined in a Slit Nanopore. \emph{J. Phys.
  Chem. C} \textbf{2011}, \emph{115}, 16544--16554\relax
\mciteBstWouldAddEndPuncttrue
\mciteSetBstMidEndSepPunct{\mcitedefaultmidpunct}
{\mcitedefaultendpunct}{\mcitedefaultseppunct}\relax
\EndOfBibitem
\bibitem[Merlet \latin{et~al.}(2013)Merlet, P\'ean, Rotenberg, Madden, Daffos,
  Taberna, Simon, and Salanne]{Merlet13d}
Merlet,~C.; P\'ean,~C.; Rotenberg,~B.; Madden,~P.~A.; Daffos,~B.;
  Taberna,~P.~L.; Simon,~P.; Salanne,~M. Highly confined ions store charge more
  efficiently in supercapacitors. \emph{Nat. Commun.} \textbf{2013}, \emph{4},
  2701\relax
\mciteBstWouldAddEndPuncttrue
\mciteSetBstMidEndSepPunct{\mcitedefaultmidpunct}
{\mcitedefaultendpunct}{\mcitedefaultseppunct}\relax
\EndOfBibitem
\bibitem[Stukowski(2010)]{OVITO}
Stukowski,~A. Visualization and analysis of atomistic simulation data with
  OVITO - the Open Visualization Tool. \emph{Modelling Simul. Mater. Sci. Eng.}
  \textbf{2010}, \emph{18}, 015012\relax
\mciteBstWouldAddEndPuncttrue
\mciteSetBstMidEndSepPunct{\mcitedefaultmidpunct}
{\mcitedefaultendpunct}{\mcitedefaultseppunct}\relax
\EndOfBibitem
\bibitem[Chmiola \latin{et~al.}(2006)Chmiola, Yushin, Gogotsi, Portet, Simon,
  and Taberna]{Chmiola06}
Chmiola,~J.; Yushin,~G.; Gogotsi,~Y.; Portet,~C.; Simon,~P.; Taberna,~P.~L.
  Anomalous Increase in Carbon Capacitance at Pore Sizes Less Than 1 Nanometer.
  \emph{Science} \textbf{2006}, \emph{313}, 1760--1763\relax
\mciteBstWouldAddEndPuncttrue
\mciteSetBstMidEndSepPunct{\mcitedefaultmidpunct}
{\mcitedefaultendpunct}{\mcitedefaultseppunct}\relax
\EndOfBibitem
\bibitem[Raymundo-Pi{\~{n}}ero \latin{et~al.}(2006)Raymundo-Pi{\~{n}}ero,
  Kierzek, Machnikowski, and B\'eguin]{Raymundo-Pinero06}
Raymundo-Pi{\~{n}}ero,~E.; Kierzek,~K.; Machnikowski,~J.; B\'eguin,~F.
  Relationship between the nanoporous texture of activated carbons and their
  capacitance properties in different electrolytes. \emph{Carbon}
  \textbf{2006}, \emph{44}, 2498 -- 2507\relax
\mciteBstWouldAddEndPuncttrue
\mciteSetBstMidEndSepPunct{\mcitedefaultmidpunct}
{\mcitedefaultendpunct}{\mcitedefaultseppunct}\relax
\EndOfBibitem
\bibitem[Lee \latin{et~al.}(2017)Lee, Barthel, Dlotko, Moosavi, Hess, and
  Smit]{Lee17}
Lee,~Y.; Barthel,~S.~D.; Dlotko,~P.; Moosavi,~S.~M.; Hess,~K.; Smit,~B.
  Quantifying similarity of pore-geometry in nanoporous materials. \emph{Nature
  Commun.} \textbf{2017}, \emph{8}, 15396\relax
\mciteBstWouldAddEndPuncttrue
\mciteSetBstMidEndSepPunct{\mcitedefaultmidpunct}
{\mcitedefaultendpunct}{\mcitedefaultseppunct}\relax
\EndOfBibitem
\end{mcitethebibliography}

\providecommand{\latin}[1]{#1}
\makeatletter
\providecommand{\doi}
  {\begingroup\let\do\@makeother\dospecials
  \catcode`\{=1 \catcode`\}=2 \doi@aux}
\providecommand{\doi@aux}[1]{\endgroup\texttt{#1}}
\makeatother
\providecommand*\mcitethebibliography{\thebibliography}
\csname @ifundefined\endcsname{endmcitethebibliography}
  {\let\endmcitethebibliography\endthebibliography}{}

\end{document}